\documentstyle[10pt,aas2pp4,epsf]{article}

\lefthead{Keohane, Rudnick \& Anderson}
\righthead{X-ray and Radio Emission from Cas~A}

\begin{document}

\title{A Comparison of X-ray and Radio Emission\\
from the Supernova Remnant Cassiopeia~A}
\author{Jonathan W. Keohane\altaffilmark{1}, Lawrence
Rudnick\altaffilmark{2} and Martha C. Anderson\altaffilmark{3}}
\affil{The Department of Astronomy, The University of Minnesota  \\
 116 Church St. SE, Minneapolis, MN \ 55455}

\authoraddr{Jonathan W. Keohane \hfill \linebreak Code 662,
NASA\discretionary{-}{}{/}GSFC \\ Greenbelt, MD \ 20771 }
\authoremail{jonathan@lheamail.gsfc.nasa.gov, larry@mazel.spa.umn.edu, 
anderson@bob.soils.wisc.edu}

\altaffiltext{1}{present address: Code 662, NASA\discretionary{-}{}{/}GSFC, Greenbelt, MD
\ 20771; \linebreak[3] E-mail: jonathan@lheamail.gsfc.nasa.gov}
\altaffiltext{2}{E-mail: larry@mazel.spa.umn.edu}
\altaffiltext{3}{present address: Dept of Soil Science, 
University of Wisconsin, 1525 Observatory Drive, Madison WI \ 53706; \\ 
E-mail:  anderson@bob.soils.wisc.edu}

\vspace {1.25 cm}
\centerline{ To appear in Volume 446 of }
\centerline {\em The Astrophysical Journal}
\centerline{ August 1, 1996 } 

\begin{abstract}
We compare the radio and soft X-ray brightness as a function of
position within the young supernova remnant Cassiopeia~A\@.  A
moderately strong correlation (r = 0.7) was found between the X-ray
emission (corrected for interstellar absorption) and radio emission,
showing that the thermal and relativistic plasmas occupy the same
volumes and are regulated by common underlying parameters. The
logarithmic slope of the relationship, $\ln(S_{\rm X\!-\!ray}) = 1.2
\times \ln(S_{\rm radio}) + \ln(k)$ implies that the variations in
brightness are primarily due to path length differences.  The X-ray and
radio emissivities are both high in the same general locations, but
their more detailed relationship is poorly constrained and  probably
shows significant scatter.

The strongest radio and X-ray absorption is found at the western
boundary of Cas~A\@.  Based on the properties of Cas~A and the 
absorbing molecular cloud, we argue that they are physically interacting.

We also compare ASCA derived column densities with $\lambda21$~cm {\sc
H~i} and $\lambda$18~cm OH optical depths in the direction of Cas~A, in
order to provide an independent estimate of ISM properties.   We derive
an average value for the {\sc H~i} spin temperature of $\approx 40
\arcdeg K$ and measure the ratio OH/H$_2$ , which is nominally larger
than previous estimates.

\end{abstract}


\pagebreak[3]

\section{Introduction}

The basic hydrodynamical structure of idealized young supernova
remnants (SNRs) seems well-under\-stood from a theoretical standpoint
(Gull 1973a, Chevalier 1982)\@.  We expect to find an outer shock, a
contact discontinuity between the shocked circumstellar medium and the 
ejecta, and a reverse shock moving into and decelerating the ejected 
material. Each of these should give rise to radio and X-ray radiation, 
with different emissivities depending on the local physical processes.

Observationally, the situation is far from this ideal.  In the best 
studied young SNR, Cas~A,  none of these structures can be clearly
identified (Anderson \& Rudnick 1995, hereafter A\&R)\@.  There are
also  questions about the nature of the outer shock, where the expected 
tangential magnetic fields are not seen (e.g., Kepler - Dickel et al.\  
1989,   and  Tycho - Dickel et al.\ 1991)\@.
Inhomogeneities on a variety of scales also complicate the
observational as well as the theoretical  pictures (e.g.\ Borkowski
et al.\ 1992, Cliffe \& Jones 1994, Jun \& Norman 1994)\@.
 
We need to clarify the nature of actual SNR structures both to
understand the hydrodynamics and also to begin addressing important
physical issues such as magnetic field amplification and relativistic
particle acceleration.  Although reasonable theoretical mechanisms
exist for these processes (Gull 1973b, Reynolds \& Ellison 1992), the
observational signatures are far from clear ({\it e.g.}\ Anderson et
al.\  1994)\@.  One fruitful approach to addressing such questions may
be a careful examination of the relationship between the X-ray and
radio emissivities within a remnant, because of the different physical
processes involved.

The bulk of the X-ray emission at low energies results from thermal
line emission (Becker et al.\ 1979, Holt et al.\ 1994, hereafter HGTN),
depending primarily on the temperature ($\approx 3$ {\rm keV}) and
density  ($\approx 10~{{\rm cm}^{-3}}$) of the plasma carrying most of
the mass and momentum (Fabian et al.\ 1980, hereafter F80)\@. On the
other hand, the radio emission is synchrotron radiation from
relativistic electrons ($\epsilon \approx 0.05-8$ GeV) in magnetic
fields of $\approx 100 \: \mu$G (Cowsik \& Sarkar 1980)\@.  In one
remnant, SN1006, X-ray synchrotron radiation is probably present at keV
energies (Koyama et al., 1995), although this is an exceptional case.

Very little quantitative work has been done on the comparison of X-ray
and radio emissivities in SNRs\@.  The canonical wisdom is that the two
are well-correlated on large scales, but show little correlation at
smaller spatial scales (F80,  Matsui et al\  1984, hereafter MLDG)\@.
MLDG studied these relations in Kepler, where they divided the remnant
into twelve sectors and found a moderate correlation of the form
$\ln\left( S_{\rm X\!-\!ray} \right) \approx (1.1 \; \mbox{to} \; 2.5)
\times \ln(S_{\rm radio})  + \ln(\mbox{scale factor}) $\@.

We chose Cas~A for study  because of the availability of both high
quality radio and X-ray data.  In the {\rm cm} wavelength range,  Cas~A
is the brightest object in the sky outside of the solar system.  At an
age of 300 years, it is believed to be in a pre-Sedov phase, and is
situated  $3.4^{+0.3}_{-0.1}$ kpc away (Reed et al.\ 1995), at the far
edge of the Perseus arm.  The column densities of hydrogen between here
and Cas~A ($N_{\rm H}\approx10^{22}~{\rm cm}^{-2}$) are such that the
optical depths of 1-2 {\rm keV} X-rays are of order unity.  Therefore,
column densities inhomogeneously distributed across the remnant, such
as due to structures local to Cas~A and the Perseus arm, can play a
large role in determining the apparent soft X-ray morphology.  For
these same reasons, Cas~A is an excellent choice as a background source
for radio and X-ray interstellar medium (ISM) studies.

The ISM has been well-studied in $\lambda 21$ cm absorption (Mebold and
Hills 1975, hereafter MH75; Bieging et al.\ 1991, hereafter BGW; and
Schwarz et al.\ 1996, hereafter SGK)\@.  Although BGW's VLA study was
the highest resolution and most detailed, it only covered the velocity
range of the Perseus arm.  SGK's Westerbork study, though only at a
resolution of 30\arcsec, covered both the Orion and Perseus spiral
arms.    In addition there have been numerous molecular absorption
studies using Cas~A --- in H$_{2}$CO (Goss et al.\ 1984), in CO
(Troland et al.\ 1985, hereafter TCH and Wilson et al.\ 1993, hereafter
WMMPO), NH$_{3}$ (Batrla et al.\ 1984 and Gaume et al.\ 1994) and OH
(Bieging \& Crutcher 1986, hereafter BC)\@.  The absorption patterns in
the various {\it molecules} are similar to each other, but very
different than that of the {\sc H~i}\@.  However, {\sc C ii} seems to
be correlated with the {\sc H~i} instead of the molecules
(Anantharamaiah et al.\ 1994)\@.

Rasmussen (1996, hereafter R96) studied the spatial dependence of X-ray
model parameters using the ASCA satellite, resulting in a total column
density ($N_{\rm H}$) map.  In this paper, we compare the radio
absorption data of BC and SGK to the $N_{\rm H}$ map of R96\@.  From
this we measure the scaling relation between column density and
equivalent line widths from the radio absorption measurements of SGK
and BC\@.  This allows measurements of the average {\sc H~i} spin
temperature and the $N_{\rm OH}/N_{\rm H_2}$ abundance of the ISM to be
calculated.

\begin{figure}
\epsscale{0.85}
\plotone{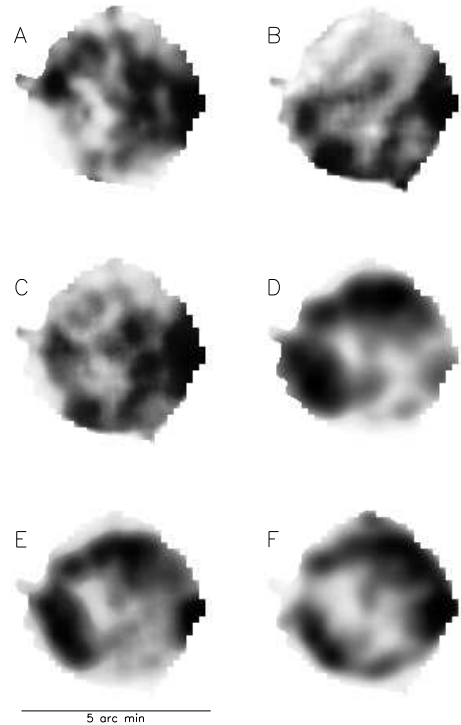} \\
\caption{30\arcsec\ resolution images of Cas~A\@. The
histogram-equalization method of scaling was used to enhance the
images.  The quantities represented are:  (A) the equivalent width of
the $\lambda21$~cm line; (B) the equivalent width of the $\lambda18$~cm
(OH) line; (C) the total column density as derived from images A and B;
(D) the logarithm of the ROSAT HRI image; (E) the logarithmic HRI image
corrected for absorption; (F) a logarithmic $\lambda20$ cm VLA
continuum map.  Histograms of the quantities shown in images A, B and C
are shown in figure~\ref{histograms}; image D ranges from -2.6 -- -0.8
$\rm \ln(cts~s^{-1}~beam^{-1})$; a plot of image E versus image F is shown in
figure~\ref{L.v.X}\@. \label{images} }
\end{figure}

\begin{figure}
\epsscale{1.00}
\plotone{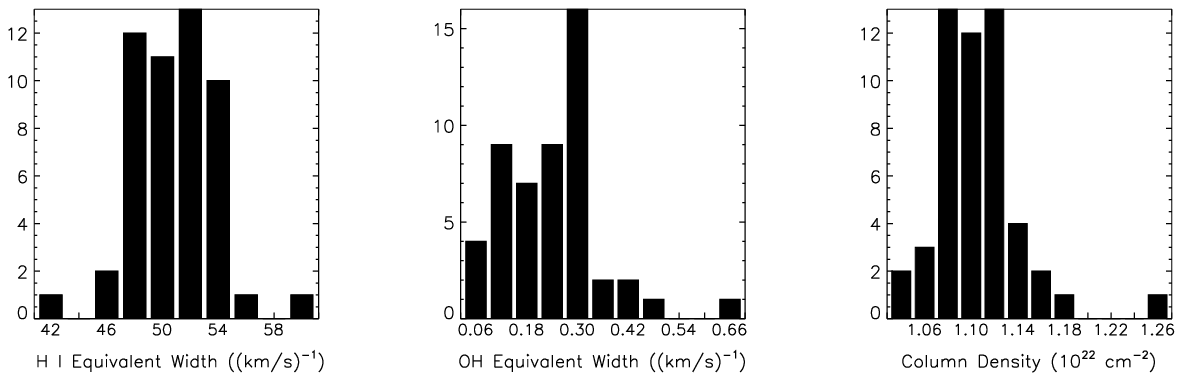} \\ 
\caption{Histograms of $\lambda21$ cm
equivalent width, $\lambda18$ cm equivalent width and the derived
 column density ($ N_{\rm H} = D \left( EW_{\sc Hi}\right) +
E\left(EW_{\rm OH}\right) + F$)\@.
\label{histograms}}
\end{figure}

\section{Analysis}

\subsection{Comparing Cas A's X-ray and Radio Surface Brightnesses 
\label{Cas_A_analysis}}

In order to compare the HRI X-ray ($S_{\rm HRI}$) and radio $(S_{\rm
radio})$ surface brightnesses, we must first correct for the soft X-ray
optical depths, which are of order unity.  The higher energy ASCA data
do not suffer from this problem, but a full analysis of those data at
sufficient resolution has not yet been done.  In order to estimate the
X-ray optical depths, we made use of BC's $\lambda18$ cm OH and SGK's
$\lambda21$ cm {\sc H~i} absorption data (see Fig.~\ref{images} A \& B
and Fig.~\ref{histograms}) to calculate the total column density
($N_{\rm H}$, Fig.~\ref{images}C and Fig.~\ref{histograms}) of hydrogen
along each line of sight to the remnant.  Once an $N_{\rm H}$ map has
been calculated, a corrected image of unabsorbed soft X-ray emission
($S_{ \rm  X_{ctd}}$, Fig.~\ref{images}E) can be derived.  Next, we
can compare the actual thermally produced X-ray surface brightness
($S_{ \rm  X_{ctd}}$) with Cas~A's well-known radio synchrotron
morphology ($S_{\rm radio}$, Fig.~\ref{images}F)\@.  We first describe
a direct method for carrying out these corrections, and then the
alternative scheme we found it necessary to use.

We start by writing the X-ray optical depth as
\begin{equation}
\tau_{_{\rm X}} \equiv \sigma_{_{\rm X}}^{_{\rm HRI}} \times N_{\rm H}
\end{equation}
where $ \sigma_{_{\rm X}}^{_{\rm HRI}}$ is the cross-section per
hydrogen atom averaged over the HRI bandpass;  $ N_{\rm H}$ is the
total number of hydrogen atoms along the line of sight.  We can then
calculate
\begin{equation}
N_{\rm H} = N_{\sc Hi} + 2 \times N_{\rm H_2}
\end{equation}
For the atomic hydrogen,
\begin{equation}
\label{eqHI}
N_{\sc Hi}   =  \left(1.83 \times 10^{18}\;
EW_{\sc Hi}\right)  \left(\frac{T_{\rm spin}}{\rm \arcdeg K}\right) 
\: \rm cm^{-2} 
\end{equation}
where $T_{\rm spin}$ is the spin temperature of the atomic hydrogen, and 
the equivalent width in {\sc H~i}  is calculated
by integrating the observed optical depths ${\tau_v}$ over all 
velocities
\begin{equation}
 EW_{\sc Hi} \equiv \int \tau_v \:{\rm d}v 
\end{equation}
In the case of the molecular hydrogen, for which direct measurements are
not available, we can use the OH absorption data, and write
\begin{equation}
\label{eqOH}
N_{\rm H_2} = \frac{N_{\rm H_2}}{N_{\rm OH}} \times \left(2.2 \times 10^{14}\;
EW_{\rm OH}\right) \: \left(\frac{T_{\rm ex}}{\rm \arcdeg K}\right) 
\: \rm cm^{-2} 
\end{equation}
with equivalent width defined as above.

This approach (of finding the X-ray absorbing column densities {\em a
priori} from radio line data) is the straight forward one, but it has
several problems because of uncertainties in the scaling constants that
must be used.  The effective X-ray cross-section has been modeled by
Morrison and McCammon (1983, hereafter M\&M), who showed that the
monoenergetic cross section depends strongly on photon energy and
changes discontinuously at quantum thresholds.  Therefore,
$\sigma_{_{\rm X}}^{_{\rm HRI}}$ is highly dependent on Cas~A's soft
X-ray spectrum, and thus has significant uncertainties.  In addition,
it is difficult to know what value to use for the {\sc H~i} spin
temperature, as was demonstrated by MH75\@.   Another major uncertainty
comes from the OH abundance value.   These issues will be discussed
more fully in section \ref{ISM_sec}\@.

We therefore adopted a different approach, simultaneously determining
the X-ray\discretionary{-}{}{/}radio relationship and the absorption corrections 
by minimizing the quantity 
\begin{equation}
\label{chi2}
\chi^2 \equiv \frac{[\ln(S_{ \rm  X_{ctd}})- \eta \ln(S_{\rm radio})
- \ln(k)]^2} {\sigma_{_{ \rm  X_{ctd}}}^2 + (\eta_{_{ i-1}} \times
\sigma_{\rm radio})^2}
\end{equation}
with respect to the parameters $\eta, A, B,$ and $[C - \ln(k)]$, where 
the logarithmic {\em corrected} X-ray image is given by:
\begin{eqnarray}
\ln(S_{ \rm  X_{ctd}}) & = & \ln(S_{\rm HRI}) + \tau_{_{\rm X}} \nonumber \\
 &  =  & \ln(S_{\rm HRI}) + A \times  EW_{\sc Hi} \label{defABC} \\ 
 &      &  + B \times  EW_{\rm OH} + C  \: .  \nonumber
\end{eqnarray}

The parameter $\eta$ measures the logarithmic scaling between the X-ray
and radio emission.  The parameter $[C - \ln(k)]$ contains both
information on the normalization of emissivities (equation \ref{chi2},
parameter $\ln(k)$), as well as allowing for absorption that is not
well-modelled by the {\sc H~i} and OH optical depths (equation
\ref{defABC}, parameter $C$)\@.  This could be due, e.g., to {\sc H~i}
saturation, variations in  $T_{\rm spin}$, which are not reflected in
X-ray absorption, and the likely existence of Orion Arm molecular gas
outside of the BC measurements.

Because the errors in the data are a function of the radio and X-ray
flux, it is important that the error images are included in the
determination of $\chi^2$ (equation \ref{chi2}). 

As a simplifying procedure, we performed the fits iteratively, holding
the errors fixed as propagated from the previous best fit parameters
($A_{i-1}$, $B_{i-1}$, and $\eta_{_{ i-1}}$):
\begin{eqnarray}
\sigma_{_{\rm X_{ctd}}}^2 & = & \sigma_{\ln(S_{\rm HRI})}^2 + 
      (A_{i-1} \times \sigma_{EW_{\sc Hi}})^2  \label{defsigx} \\
 & &  +   (B_{i-1} \times  \sigma_{EW_{\rm OH}})^2 \: . \nonumber
\end{eqnarray}

Prior to performing the $\chi^2$ analysis, the maps were prepared as
follows.  In order to match the resolution of the {\sc H~i} data, we
smoothed the OH line and continuum maps to a resolution of 30\arcsec,
before calculating the optical depths and equivalent widths.  We also
smoothed the epoch 1990 (3.4 ks live-time) ROSAT HRI image (obtained
from the HEASARC public archive) and the epoch 1987 $\lambda20$ cm VLA
continuum map described by A\&R, to 30\arcsec\ resolution. Errors in
each map were calculated using standard methods.  The resulting maps
contain 51 independent beams, and are shown in figure~\ref{images}\@.
Histograms of the radio absorption values are shown in
figure~\ref{histograms}\@.

\begin{figure}
\epsscale{1.0}
\plotone{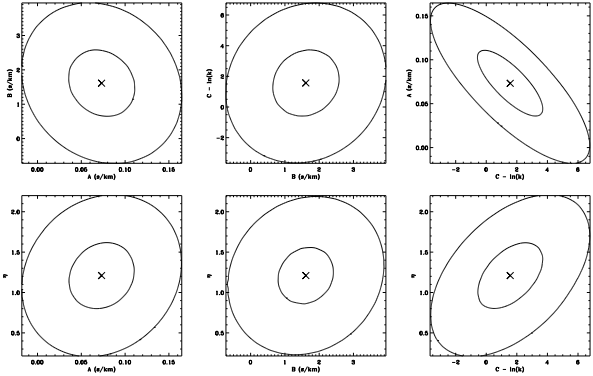} \\ 
\caption{70\% and 99\% confidence levels derived from the goodness of fit
$\chi^2$ obtained by fitting a power-law relation between the
radio and corrected X-ray morphological distributions.  They are shown
here as a function of the parameters $A$, $B$, $[C - \ln(k)]$
and $\eta$ as defined in the text\@.  The $\times$ in each plot represents the minimum $\chi^2$\@.  \label{AB_ctrs}}
\end{figure}

The minimum value of $\frac{\chi^2}{N}$ was 47, showing that there is a
large amount of scatter still unaccounted for in the
X-ray\discretionary{-}{}{/}radio relation.  The number of degrees of
freedom, $N$, coincidentally, also equals 47.  In order to calculate
70\% and 99\% confidence limits ($\delta_{X70},\delta_{X99}$) for
parameter $X$, we determined the value of \begin{equation} \chi^2(X \pm
\delta_{X70,X99}) \equiv \chi^2_{\rm min} \times (1.097, 1.542)
\end{equation} where $\chi^2_{\rm min}$ is the global minimum value of
$\chi^2$\@.  Following Bevington \& Robinson (1992), while parameter $X$
is being varied, all other parameters are allowed to float to minimize
$\chi^2$ for that value of $X$\@. Figure~\ref{AB_ctrs} shows contour
plots of the confidence levels, as a function of parameters $A$ and
$B$, $\eta$ and $[C - \ln(k)]$\@.  

\begin{figure}
\epsscale{0.75}
\plotone{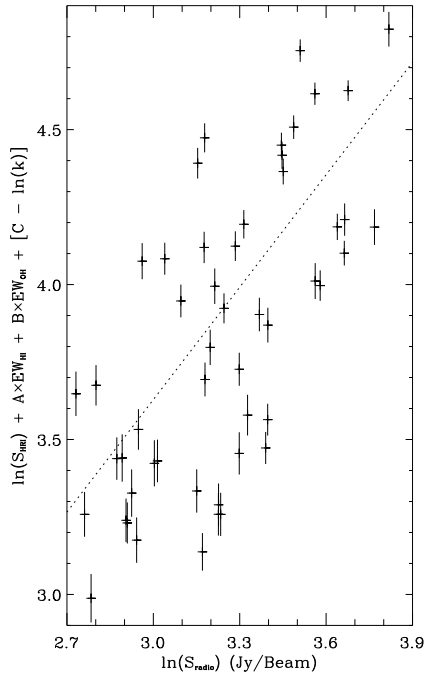}
\caption{Plot of the logarithmic
soft X-ray emission corrected for absorption using the best fit values
for parameters $A$, $B$, and $[C - \ln(k)]$ as defined in
equation~\ref{defABC} versus the logarithmic $\lambda20$ cm
emission\@.  The line represents our the best fit model:  $\ln(S_{
\rm  X_{ctd}}) - \ln(k) = \eta \ln(S_{\rm radio})$\@.
\label{L.v.X}}
\end{figure}

Figure \ref{L.v.X} shows the correlation between the thermal X-ray
emission $\ln(S_{ \rm X_{ctd}})$ and radio synchrotron
radiation ($\ln(S_{\rm radio})$)\@.  The best fit slope is $\eta = 1.21
\pm 0.41$, (70\% confidence limit), with the error probably dominated
by real scatter in the X-ray\discretionary{-}{}{/}radio relation.  In
section \ref{disc}, we will discuss the competing effects of emissivity
and path length that can contribute to such differences.

\subsection{Interstellar Medium Parameters}
\label{ISM_sec}
In this section, we introduce a technique for determining physical
parameters of the ISM.  Although our results have large uncertainties,
the method is useful in confirming previous estimates, and could be
extended to higher accuracy.  Using the above procedure, we
simultaneously determined  the scaling parameters between {\sc H~i} and
OH optical depths and soft X-ray absorption due to the line-of-sight
ISM column density.   The best fit values are $A = 0.073 \pm 0.039$
$({\rm km/s})^{-1}$, $B = 1.6 \pm 1.0$ $({\rm km/s})^{-1}$ and $[C -
\ln(k)] = 1.6 \pm 2.2$ (Fig.~\ref{AB_ctrs})\@.

 The X-ray absorption and the total column density can also be measured
more directly through X-ray spectral fitting.  With spectral fitting,
the effective X-ray cross-section, which changes strongly as a function
of photon energy (M\&M), can be explicitly included.  R96 assumed
Cas~A's X-ray spectrum to consist of power-law emission plus many
Gaussian spectral emission lines (see HGTN), absorbed by the ISM
assuming M\&M's effective cross-sections\@.   This resulted in a $32
\times 32$ map (19 \arcsec/pixel) for each parameter fit, including the
column density.  A factor of 2 oversampling was used, so the ``beam
size'' is approximately 40\arcsec; this is also a function of ASCA's
complex point spread function (Tanaka et al., 1994).  However, since
these fits did not cover the whole image of Cas A, and because they
rely on a ``super-resolution'' of the data, we chose not to use them for
the original correction of the X-ray maps.  We will, however, use the
ASCA data to derive parameters of the interstellar medium.

In this section, we compare our optical depth image to R96's ASCA $N_{\rm
H}$ image to derive a second set of scaling parameters for $EW_{\sc
Hi}$ and $EW_{\rm OH}$\@. Combining this with the previously determined
scaling parameters, and using equations \ref{eqHI} and \ref{eqOH}, we
can then determine both the effective ROSAT HRI cross section as well
as some important ISM parameters.

To perform this analysis, we first aligned the ASCA field with our
radio images by maximizing the cross-correlation of R96's $N_{\rm
H_{\rm ASCA}}$ image with our optical depth image ($A \times EW_{\sc
Hi} + B \times EW_{\rm OH}$)\@.  This yielded a total of 25 independent
30\arcsec\ beams with which to compare the X-ray and radio absorption.  With
$N_{\rm H_{\rm ASCA}}$ in real physical units ($10^{22} {\rm cm}^{-2}$)
and the equivalent widths in  km/s, we then minimized the quantity:
\begin{equation}
\chi^2(D,E,F) \equiv \sum {\left(N_{\rm H_{\rm ASCA}} - 
                                 N_{\rm H_{Radio}} \right)^2}
\end{equation}
where D, E, and F are defined by
\begin{equation}
\label{defDEF}
 N_{\rm H_{Radio}} =
D \left( EW_{\sc Hi}\right) + E \left( EW_{\rm OH} \right) + F
\end{equation}

\begin{figure}
\epsscale{1.0}
\plotone{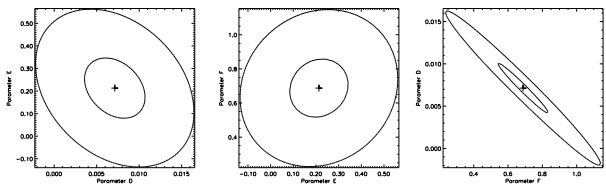} \\
\caption{The $\chi ^2$
between the ASCA  and radio absorption line derived column densities as
a function of the scaling parameters D, E and F, which are defined such
that $ N_{H} = D \left( EW_{\sc Hi}\right) +
E\left(EW_{\rm OH}\right) + F$\@.  At each point in the plots, the parameter not shown was set to the value which would minimize $\chi^2$ there.  Contour levels
represent confidence levels of 70\% and 99\%\@. \label{DEF}}
\end{figure}

The best fit values for parameters $D$,$E$ and $F$ are:
$(7.1 \pm 3.8) \times 10^{-3}$, $0.213 \pm  0.141$, and $ 0.69 \pm
0.19$ respectively (70\% confidence); the $\chi^{2}$ dependence on
$D$,$E$ and $F$ is shown in Figure~\ref{DEF}\@.   The confidence limits
are defined as:
\begin{equation}
\chi^2 (X \pm \delta_{X70,X99}) \equiv \left( 1.134, 1.831 \right)
\times \chi^2_{\rm min}
\end{equation}
as appropriate for 22 degrees of freedom.  The use of $\chi^2_{\rm min}$ is necessary because R96 did not provide an error image.

At this point, we have two sets of scaling parameters for $EW_{\sc Hi}$
and $EW_{\rm OH}$\@.  One set depends on the HRI cross section;  the
other does not.  We can therefore use equations \ref{defABC} and
\ref{defDEF} to find that \begin{equation} \sigma_{_{\rm X}}^{_{\rm
HRI}} = \frac{A}{D} = \frac{B}{E} \end{equation} This yields
$\frac{A}{D} = 10.3 \pm 7.7 $ and $ \frac{B}{E} = 7.6 \pm 6.8$ with a
weighted average of $\sigma_{_{\rm X}}^{_{\rm HRI}} = (8.8 \pm 5.1)$; all
are in units of  $10^{-22} {\rm cm}^{2}$\@.  This value of $\sigma_{_{
\rm X}}^{_{\rm HRI}}$ is several times larger than the rough estimate we
made by averaging the M\&M cross-section values over
the bandpass, weighted by the HRI effective area, and assuming a flat
spectrum for Cas~A\@. If this discrepancy is real, then it indicates
that on average, Cas~A's source spectrum is probably decreasing with
increasing energy in the HRI bandpass.

From our measured parameters $D$ and $E$, and assuming that we have
isolated the contributions due to {\sc H~i} and OH in our $\chi^2$
analysis, we can then use equations \ref{eqHI} and \ref{eqOH} and the
values of $D$ and $E$ to calculate $T_{\rm spin}$ and $N_{\rm
OH}/N_{\rm H_2}$\@. However, the presence of a significant non-zero value
for $F$ shows that we have not successfully modelled all of the X-ray
absorption, and we will have to take this into account.

First, we consider the molecular hydrogen component. Figure~\ref{DEF} 
shows that there is very little correlation between
parameter $E$ and either parameter $D$ or $F$\@.  This implies that our
best fit parameter $E$ is robust to uncertain contributions to the
absorption from other ISM components\@.  We therefore find the fraction
of OH in the Perseus spiral arm as: 
\begin{equation}
\frac{N_{\rm OH}}{N_{\rm H_{2}}} = \left( 4.1 \pm 2.7
\right) \times 10^{-6} \, \left( \frac{T_{\rm ex}}{20 \rm  \arcdeg K}
\right) \: . 
\end{equation}
This derived ratio is nominally higher than the $(2.9 \pm 2.7) \times
10^{-7}$ quoted by WMMPO.

The contribution of the atomic component (parameter $D$) is more
problematic, because of its correlation with parameter $F$ (see
Fig.~\ref{DEF})\@.  Formally, we can calculate an average spin
temperature for the {\sc H~i} as \begin{equation} T_{\rm spin} =
D/(1.83 \times 10^{-4}) = (39 \pm 21) \, {\rm \arcdeg K} \: .
\end{equation} This temperature is the same as the spin temperature
derived by Payne et al.\ (1994) towards Cas~A and in the range of those
found by Kalberla, et al. (1985), toward 3C147\@. The major
contribution to the error in $D$ actually comes from its correlation
with $F$\@.

There are three likely sources for a non-zero $F$:  Orion arm molecular
gas (not measured by BC), saturated {\sc H~i} and hot intercloud {\sc
H~i}\@.

TCH observed Orion arm $^{13}$CO column densities to be approximately
20\% of the Perseus arm column density.  This implies that only about
$6 \times 10^{20} \: {\rm cm}^{2}$ of $N_{\rm H_{\rm ASCA}}$ ($\approx
10\%$ of parameter $F$) can be accounted for by molecular gas outside of
BC's velocity range.

BGW named the highly saturated {\sc H~i} absorption feature near a
velocity of -48 km/s ``the curtain'', because of its spatial uniformity
and high optical depth (mostly above 5)\@.   Here SGK's {\sc H~i}
equivalent width measurements underestimate the actual absorption.
If SGK accounted for about half of the {\sc H~i} in ``the
curtain'', the unmeasured portion should account for about $1$-$2
\times 10^{21} \: {\rm cm}^{2}$ of $N_{\rm H_{\rm ASCA}}$ or $\approx
15\%$ of parameter $F$)\@.

MH75 used the Effelsberg 100m telescope to observe the $\lambda 21$ cm
line in both emission and absorption toward Cas~A\@.  Their
observations are consistent with a two-temperature {\sc H~i} model;
where the hot component has spin temperatures in excess of about 3500
\arcdeg K, which would be observed only minimally by SGK (equation
\ref{eqHI})\@.  MH75's estimated cloud/intercloud mass ratio is about 1:2,
which could account for all of our measured parameter $F$\@.

These factors complicate our determination and interpretation of
$T_{spin}$, as described above, so it should be considered simply as a
characteristic value for the cooler gas.

\section {Discussion}
\label{disc}  

\subsection{The X-ray/Radio Emissivity Relation}

   The long-term objective of comparing the X-ray and radio emission in
Cas A is to identify the state and structure of the thermal and
relativistic plasmas and the physical connections between them.  The
X-ray emission spectrum from Cas~A is itself believed to come from two
different plasmas (Jansen et al.\ 1988)\@.  The low temperature
plasma ($kT< 1~{\rm keV}$) is most likely reverse-shocked ejecta; models of the emission in the ROSAT band show that the
emissivity is a mixture of thermal bremsstrahlung and line emission. The higher temperature plasma  ($kT \approx 3~{\rm keV}$) is most likely
circumstellar matter shocked by the blast wave.  The reverse shock in
Cas~A is believed to currently dominate the soft X-ray emission (F80,
Jansen et al.); the factors that influence the relative
luminosity of these two shocks are discussed in detail by Masai
(1994)\@.  We have shown here the importance of absorption for the
low energy X-rays;  analyses based on hardness ratios (e.g., HGTN) must first correct for this effect\@.      

In the radio band, most of the emission comes from the bright ring, which is
identified with either the reverse shock, the contact discontinuity, or
both (Gull 1973 a, b)\@.  There is also  a lower surface brightness 
radio plateau beyond the ring and a wealth of structures on smaller 
scales, including knots, bow shocks, filaments, etc, with lifetimes of 
order 30 years.

To interpret the observed radio\discretionary{-}{}{/}X-ray
correlation,  we must now distinguish between surface brightness and
emissivity.  If the emissivities in a remnant do not vary spatially,
then all variations in surface brightness must be due to variations in
path length (or filling factor), and the X-ray and radio brightnesses
should be proportional to each other (logarithmic slope of 1)\@.  Some
of the brightness variations in Cas~A are clearly due to path length
differences, such as the bright ring itself.  However, even within the
bright ring, the local radio brightness may largely be due to the path
length through the emitting material at that position, rather than
large variations in the intrinsic emissivity.  The observations of Reed
et al.\ (1995) show that the apparently complete Cas~A shell is very
non-uniform in optical line emission.  The analysis of A\&R show the
presence of major dynamical asymmetries which are probably coupled to
spatial variations.  The large velocity gradients in the X-ray data
from ASCA (HGTN) also demand large-scale non-spherical structures in
the X-ray emitting material.

Given the major role played by path length variations through the
emitting material, our observed logarithmic slope of 1.2 for the
X-ray\discretionary{-}{}{/}radio {\it surface brightness} relationship
can be understood only as a lower limit to the slope relating the
actual {\it emissivities}.  If the variations in emissivity are much
less than the variations in path length, the emissivity slope could be
much higher.  We are therefore unable to comment on how the various
processes leading to soft X-ray and radio emission vary with one
another.  The same problem might easily affect the results of MLDG,
whose X-ray\discretionary{-}{}{/}radio logarithmic slope could be as
low as 1.1.  At present, it is not clear how to isolate true variations
in emissivity.

The modelling of the radio\discretionary{-}{}{/}X-ray emissivity
relationship in an inhomogeneous rapidly-evolving remnant such as Cas~A
is also quite uncertain\@.  It is important to avoid the simple scaling
relations based on X-ray thermal bremsstrahlung, such as discussed by
MLDG, because such relations ignore the dependence of line emission on
factors other than density.  In addition, we know from recent numerical
simulations that the magnetic field amplification (Jun \& Norman 1994)
and relativistic particle acceleration (Jones et al.\ 1994) reflect the
history of the plasma and cannot be simply described by current state
parameters such as density and temperature. Jun (1995) has modeled the
radio synchrotron radiation and the thermal bremsstrahlung component of
X-ray emission in his 3D MHD simulations of young supernova remnants.
Although he finds the same major features in the X-ray and radio
emission, e.g., the clumpy bright ring, there is only a weak
correlation between the two (Jun, 1996)\@.  This is due
to the strong dependence of only the radio emission on the local
magnetic field. Therefore, in order to effectively use such
observations as presented here and in MLDG, we need both more
sophisticated time-dependent analyses of the X-ray radiative transfer
and an understanding of the relationship on various scales between the
magnetic field and other hydrodynamical parameters.

\subsection{The Western Molecular Cloud}

The western edge of Cas~A is unusual in a number of ways, suggesting an interaction between the expanding SNR and a local molecular cloud. Considering the properties of Cas~A itself, we first note that the brightest radio and X-ray emission is found in this region.  HGTN show that the western region has significantly lower X-ray equivalent widths than the other major emission regions.  They interpret this as due to relatively stronger emission from the outer shock, which would be expected if it were moving into an area of higher density.  A\&R found that the motions of bright radio knots in the west showed extreme departures from the relatively uniform expansion seen in other regions.  Many western knots are actually moving back toward the center of expansion - these must be due to an external interaction.  This is also a region of steep radio spectral indices, implying that the conditions for relativistic particle acceleration are different here (Anderson \& Rudnick, 1996)\@.  

On the opposite side of the remnant, there is a break in the shell, and
groups of fast moving radio (A\&R) and optical (Fesen et al., 1988)
knots.  This so-called ``jet'' could result from expansion into a lower
density region opposite the molecular cloud.  A similar situation is
modelled by Tenorio-Tagle et al.\ (1985)\@.

Turning now to the external material, figure~\ref{images} (image C) is
the derived $N_{\rm H}$ map, which shows strong absorption on the
western side of Cas~A\@.  This dense cloud also shows up as the extreme
value in the column density histogram in Fig. \ref{histograms}.  The
cloud is at an LSR velocity of $\approx$ -40 km/s, placing it in the
Perseus arm (Batrla et al.\ 1984, Goss et al.\ 1984, TCH, BC and
WMMPO)\@.  The OH column density maps of BC around -40 km/s trace out the
regions of anomalous radio knot velocities and steep spectral indices
discussed above. A trace of this cloud may also be visible in the {\sc
H~i} measurements of BGW and SGK\@.

Gaume et al.\ (1994) studied the NH$_3$  and CO absorption
towards the bright western region. For the -39 km/s cloud, they
determined the density to be $n_{\rm H_2} \approx 1000\: {\rm
cm}^{-3}$  characteristic of dark dust clouds, but found a higher than
average kinetic temperature ($\approx 18$ \arcdeg K) and line width
($3.5$ km/s)\@.  They suggested that the high temperature could be due
to either an increased level of cosmic rays, or by cloud-cloud
collisions.  On the basis of both the unusual properties of Cas~A in
the west and the unusual cloud conditions there, it thus seems quite
likely that an interaction is currently in progress.

Wilson and Mauersberger (1994) pointed out that the circular shape of
Cas~A argues for a lack of external interaction. However, the bright
radio ring illuminates material that has just recently been
decelerated, because it is found in the same area as the optical fast
moving knots, which are travelling several times faster (A\&R)\@.   R.
Dohm-Palmer \& T.W. Jones (1996) have performed 2D numerical
simulations of an SNR expanding into a sharp ISM gradient.  They find
that at the time when the reverse shock on the high density side has
reversed the velocity of some compact features, the overall ring
deviates from being circular only by $\approx 10\%$\@.  This is the
same degree of non-circularity observed for Cas~A's bright ring, and so
removes this objection to an external interaction.

\section{Conclusion}
In this paper we have presented a technique to correct for spatially
inhomogeneous absorption of soft X-rays in Cas~A using radio absorption
data.  We find a good correlation between the soft X-ray and radio
synchrotron emission from Cas~A, but with significant scatter.  The
correlation is probably dominated by variations in path length,
implying that the X-ray and radio emissions both occupy the same
volumes.  However, we have no evidence for a more detailed relation
between their emissivities.  A quantitative interpretation of these
results requires more sophisticated modeling of both the X-ray
radiative transfer and the relativistic plasma evolution in young
SNRs.

Future X-ray\discretionary{-}{}{/}radio comparisons of Cas~A should
concentrate in at least the following two directions:  studies at
higher spatial resolution with deeper HRI measurements and comparisons
with ASCA's spatially resolved spectroscopy. With a deeper ROSAT HRI
observation, it may be possible to separate emissivity from path length
variations.  In addition, X-ray proper motions could be measured and
compared with the radio proper motions. Since HGTN's paper, Cas~A has
been used as a calibrator for ASCA, so more ASCA data have been
obtained and ASCA's response functions have been refined.  This will
enable image reconstruction techniques to be applied to narrower
bandpass images and better quality spatially resolved spectral
fitting.  Studying the correlations between the radio and X-ray
morphologies as a function of X-ray energy will allow the different
emission mechanisms and temperature and metalicity structures to be
distinguished.

We have shown that Cas A is likely to be interacting with a dense cloud
in the west.  This has affected both the properties of the remnant and
the cloud.  Such interactions may play an important role both in SNR
dynamics, and in the transfer of energy into the ISM.

We have also demonstrated a new technique for probing the ISM\@.  By
comparing X-ray and radio spectroscopic absorption measurements, the
{\sc H~i} spin temperature and molecular abundances ratios were
measured.  Future studies of other radio and X-ray bright extended
objects can significantly enhance our understanding of the ISM, by
comparing spatially resolved column densities from either the ROSAT
PSPC or ASCA with radio and far IR atomic and molecular line data.

\acknowledgments
This work was supported in part by the National Science Foundation,
through grant number AST 93-18959 to the University of Minnesota.
Funding was also provided by NASA\discretionary{-}{}{/}GSFC's
Laboratory for High Energy Astrophysics through the Graduate Student
Researchers Program.  This research has made use of data obtained from
the US ROSAT Public Data Archive which is jointly managed by the ROSAT
Science Data Center and the HEASARC\@. The HEASARC is a collaboration
of the Laboratory for High Energy Astrophysics and the NSSDC at
NASA\discretionary{-}{}{/}GSFC\@.  The following data were kindly
provided for our analysis --- OH absorption data, John Bieging; ASCA
column densities, Andrew P. Rasmussen; Westerbork {\sc H~i} absorption
data, W. Miller Goss. We are also grateful for Goss's critical
contributions as referee of this paper, and to John Dickey, B.I. Jun, Robert Petre and Ulrich Schwarz for helpful discussions and comments.

\pagebreak[3]

\end{document}